\documentclass[runningheads]{llncs}

\usepackage{graphicx}
\usepackage{hyperref}
\usepackage{commath}
\usepackage{tabularx}
\usepackage{enumitem}
\usepackage{url}
\usepackage{color}
\usepackage{xcolor}

\usepackage{amssymb}
\usepackage{csquotes}

\usepackage{multicol}
\usepackage{caption}
\usepackage{comment}

\usepackage{xspace}

\newcommand{\sysname}{GCNetOmaly\xspace}

\title{Detecting Anomalous Network Communication Patterns Using Graph Convolutional Networks}

\titlerunning{Detecting Anomalous Network Communication Patterns Using GCNs}

\begin{document}

\author{Yizhak Vaisman, Gilad Katz, Yuval Elovici, Asaf Shabtai}
\institute{Ben-Gurion University of the Negev}

\authorrunning{}

\maketitle

\begin{abstract}
To protect an organizations' endpoints from sophisticated cyberattacks, advanced detection methods are required.
In this research, we present \sysname --- a graph convolutional network (GCN)-based variational autoencoder (VAE) anomaly detector trained on data that include connection events among internal and external machines.
As input, the proposed GCN-based VAE model receives two matrices: \textit{(i)} the normalized adjacency matrix, which represents the connections among the machines, and \textit{(ii)} the feature matrix, which includes various features (demographic, statistical, process-related, and Node2vec structural features) that are used to profile the individual nodes/machines. 
After training the model on data collected for a predefined time window, the model is applied on the same data; the reconstruction score obtained by the model for a  given machine then serves as the machine's anomaly score.
\sysname was evaluated on \textit{real, large-scale} data logged by Carbon Black EDR from a large financial organization's automated teller machines (ATMs) as well as communication with Active Directory (AD) servers in two setups: unsupervised and supervised. 
The results of our evaluation demonstrate \sysname's effectiveness in detecting anomalous behavior of machines on unsupervised data. 
\end{abstract}


\section{\label{sec:introduction}Introduction}
Currently, many organizations permit Internet access, thus allowing data to be transferred to and from the organization's network.
However, the internal network's access to the Internet exposes the internal endpoints and users to a variety of threats. 
To mitigate these risks, organizations often employ countermeasures such as Web Proxy servers, antivirus software, intrusion detection systems, and endpoint detection and response (EDR). 
In most cases, these tools are based on a rule- and signature-based techniques for the identification of malicious behavior.
While such an approach is useful to issue alerts on malicious activity in real time with zero false positives, it is limited to the detection of known attacks.

Since today's attacks are highly sophisticated, advanced countermeasures are required.
Consequently, previous research explored the application of machine learning (ML) for the analysis of cybersecurity-related data, specifically network communication, in order to detect intrusion attempts and anomalies. 
These studies utilized traditional ML models~\cite{resende2018survey,horng2011novel,moustafa2019holistic}, as well as deep learning models~\cite{wu2020network,alauthman2020efficient}, and used datasets that include labeled attacks to train the models. 
Other research took an unsupervised approach (i.e., no labeled data is available for model training), for example, using k-means clustering and distance-based anomaly detection~\cite{ahmed2016survey} and LSTM~\cite{bontemps2016collective} model. 

The availability of datasets that include both normal and abnormal attack instances is limited; therefore, in this research we took an unsupervised approach for anomaly detection. 
Our proposed framework --- \sysname  --- is based on a graph convolutional network (GCN) with variational autoencoder (VAE) anomaly detector trained on data that includes connection events among internal machines (i.e., machines that are part of the organizational network) 
and external machines (i.e., servers and services on the Internet).
We used a GCN-based VAE for the detection of anomalous communication patterns because: (1) the monitored machines and the communication between them can be modeled as a graph, (2) advanced deep learning techniques for modeling graph structures are readily available, and in many domains, their popularity is increasing, and (3) when modeling the communication, it is important to consider attributes that characterize the nodes in the graph (i.e., machines) and their behavior.

As input, the proposed GCN-based VAE model receives two matrices: \textit{(i)} the normalized adjacency matrix, which represents the connections among the machines, and \textit{(ii)} the feature matrix, which includes various features (demographic, statistical, and Node2vec structural features) that are used to represent (profile) the individual nodes/machines.
After training the model on data collected for a predefined time window, the model is applied on the same data; the reconstruction score obtained by the model for a  given machine then serves as the machine's anomaly score.

\sysname was evaluated on real data logged by Carbon Black EDR from a large financial organization's automated teller machines (ATMs) and Active Directory servers (ADs). 
\sysname raised alerts on the anomalous behavior of at most five machines (ATM or machine in which communicated with AD) per day for the organization's several thousand ATMs and machine that communicated with AD servers, which is a reasonable and practical number of alerts to be handled on a single day.   
When manually inspected by security analysts, all of the ATM alerts were labeled as anomalous by the analysts, however, 15\% of them (three ATMs) were labeled as requiring no further investigation, because they were raised for a development ATM (which as expected, exhibit anomalous behavior). 
In addition, 64.71\% (11 out of 17) of the AD anomalous machine was labeled as good anomalies while the others were labeled as false positive by the analysts.
These results demonstrate \sysname's effectiveness in detecting anomalous behavior of ATM machines and AD communications on unsupervised data.

The main contributions of this work are as follows:
(1) a generic method, based on GCNs, for representing the communication patterns of the machines of an organization, and (2) an effective framework for the identification of anomalous behavior in an unsupervised setting, specifically, unlabeled network communication data.

\section{Proposed Method} \label{sec:system} 
\subsection{Overview}
\begin{figure}[h]
\centering
\includegraphics[width=\textwidth]{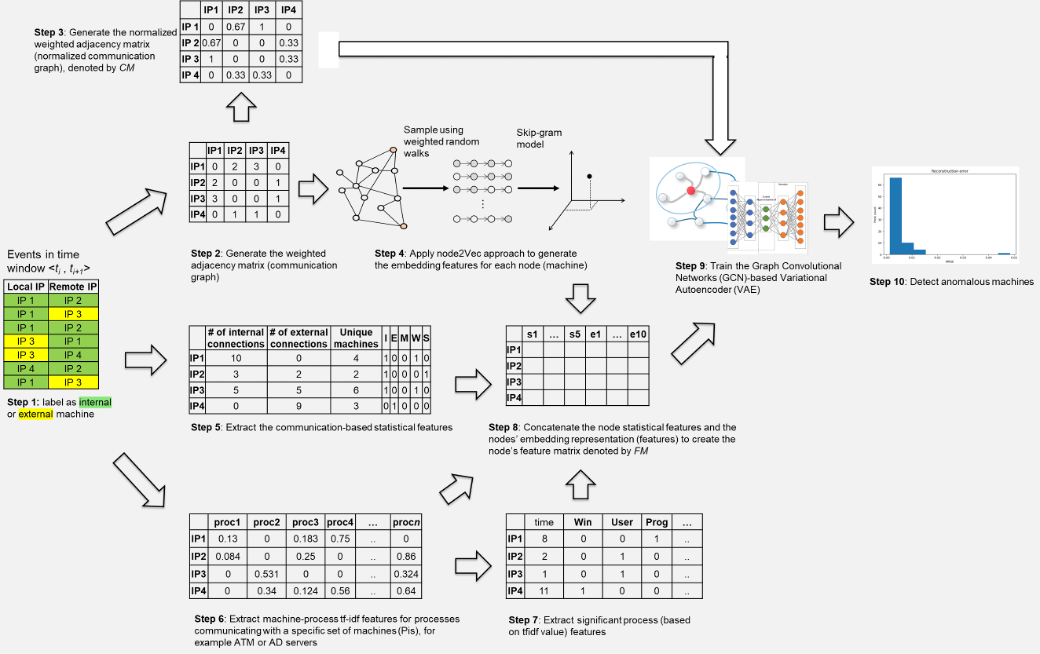}
\caption{The ten steps of the \sysname framework.}
\label{fig:framework}
\end{figure}

The proposed framework consists of four main phases (see Figure~\ref{fig:framework}):

\begin{enumerate}
\item \emph{Data collection and preprocessing} - The input to \sysname is a log containing communication data collected by the EDR agent.
In the data preprocessing phase, the raw data is cleaned and transformed prior to its processing and analysis.
In this phase, each machine is labeled as either internal or external (Step 1 in Fig.~\ref{fig:framework}).
Finally, a weighted adjacency matrix that represents the inter-machine communication is generated (Step 2) and being normalized using min-max normalization (Step 3).
    
\item \emph{Feature extraction} - Based on the preprocessed data and the weighted adjacency matrix, four types of features are extracted for each machine: node embedding features (Step 4), communication-based statistical features (Step 5), process features (Step 6), and significant process features (Step 7).
        
\item \emph{Training the GCN-based model} - Given a normalized weighted adjacency matrix (Step 3) and the extracted features (Step 8), in this phase, a GCN-based VAE anomaly detection model is trained. 

    
\item \emph{Detecting anomalous machines} - Using the reconstruction error provided by the trained model, \sysname labels a machine anomalous or normal~(step~10).
\end{enumerate}

\subsection{Notations}
We use the following notations throughout this study:
\begin{itemize}
    \item $w$ - the size of the time window for which we generate the communication graph representation and compute the features. 
    In our evaluation we set $w = 1$ day. 
    Our anomaly detection process is applied to each time window separately.
    
    \item $m \in M$ - a machine that is either monitored by the EDR agent, or communicates with such a machine ($M$ is the set of all machines processed by the model).
    
    
    \item $e = (m_i,m_j)$ 
    - a communication event between $m_i$ and $m_j$ based on the~EDR~logs.
    
    \item $c_{i,j} = Weight(m_i,m_j),  m_i,m_j \in M$ - the weight of edge $e = (m_i,m_j)$, which is computed as the number of communication events between $m_i$ and $m_j$ that were logged in the time window.
    
    \item $G_t$ - a graph $G_t = (M,E,C)$, where $M$ is the set of machines that appeared in the event logs in time window $t$, $E$ is the set of edges where $e \in E$, and $C$ are the edge's weights in time window $w$ where $c \in C$.
    
    \item $tr$ - the threshold for which a machine $m$ is considered anomalous.

\end{itemize}

\subsection{Data Preprocessing}
In the first phase, \sysname extracts the required data logged by the EDR agents installed on the organization's machines.
This data includes events that relate to network connections observed on the monitored machine.
Each event is defined by the following tuple:\\
$<id(m_i),timestamp,md5, PID, SrcIP(m_i),DestIP(m_j)>$, where:
\begin{itemize}
\item $id$ is the unique identifier of the monitored machine, 
\item $timestamp$ is the time of the network connection event, 
\item $md5$ is the md5 hash of the process that performed the communication (serves as the unique identifier of the process), 
\item $PID$ is the identifier assigned to the process; the PID may change in each execution of the process on the machine, 
\item $SrcIP$ is the IP address of the source machine, and 
\item $DstIP$ is the IP address of the target machine.
\end{itemize}

Processing and analyzing the data of $all$ machines belonging to a large enterprise's machines is a challenging task, mainly for the following two reasons:
First, the large amount of data can significantly increase the computational resources required and the processing time, especially when using deep learning models like GCNs.
Secondly, the fact that different machines and users (e.g., a developer's machine, the machine of an executive user, an active directory server, or an ATM) perform different tasks, results in high variability in the machines' network behavior (i.e., communication patterns).

For example, while on a developer's machine we can expect diverse connections to external Internet services (e.g., Google search, Stack Overflow, or GitHub), on an ATM machine we can expect that most connections will be performed with the same internal and external machines and the same internal servers.
Because of this high variability, a ML model trained on the data of all machines will be both less accurate and highly prone to false alarms.
Therefore, in the preprocessing phase, \sysname filters the data of a preselected homogeneous subset of machines sharing the same properties and having similar communication patterns; \sysname profiles the network behavior of this subset of machines, enabling it to identify machines that behave anomalously compared to the behavior of the other machines in the group.
We denote this selected subset of machines by $M' \subset M$.

To filter communication that belongs to just a subset of homogeneous machines $M'$, we consider all events in which a machine $m_i \in M'$ appears either in the source IP or the destination IP of the connection event 
and filtering out all the other events.
Finally, for each event in the filtered set, we label the machines in the source IP and destination IP as `Internal' or `External' where an internal machine is a machine that is part of the organizational network and an external machine is a machine that is outside the organization network (i.e., the Internet).

\subsection{Feature Extraction}
\sysname extracts five main types of features: communication graph, node (machine) embedding features, communication-based statistical features, process features, and significant process features.

\begin{enumerate}
    \item \emph{Communication graph.}
The weighted adjacency matrix (Step 3 in Figure~\ref{fig:framework}) of a graph is used to present communication among the machines as a sparse matrix. 
The matrix size is $|M| \times |M|$, where $|M|$ is the number of machines (identified by the IP addresses) observed within the selected time window $t$.
The value of cell $(i,j)$ in the matrix is $c_{i,j}$, which represents the number of communication events between $m_i$ and $m_j$ within the selected time window.
The normalized weighted adjacency matrix (Step 3) is a min-max normalization of the weighted adjacency matrix in which the value of cell $(i,j)$ is computed as:
$M[i,j] = (c(i,j) - min(C))/(max(C) - min(C))$. 

\item \emph{Communication-based statistical features.}
For each $m \in M$ we extract the following five statistical features based on the communication events logged within the selected time window $t$ (Step 5 in Figure~\ref{fig:framework}):

\begin{itemize}
    \item \textit{internal\_communications -} \# of communication events in which $m$ communicates with an internal machine (i.e., organization internal IP address) 
    ($m$ can be the source or destination).
    \item \textit{uni\_outgoing\_machines -} \# of unique machines with which $m$ communicated.
    \item \textit{uni\_incoming\_machines -} \# of unique machines that communicated with~$m$.
    \item \textit{machine\_type -} one-hot vector for $m$ with five values: M for mobile, S for server, W for workstation, I for internal unknown type, and E for external.
    \item \textit{rare\_processes -} \# of processes that used only by $m$ on time window $t$.
\end{itemize}

Note that all features (except the machine\_type) are normalized using min-max normalization.

\item \emph{Node embedding features.}
Using this set of features, we attempt to capture the topological representation of $m$ within the overall communication graph generated for time window $t$.
This is performed as follows:
First, we represent the network connections among the monitored machines as an \textit{undirected weighted} graph (using the weighted adjacency matrix shown in Step 2 in Figure~\ref{fig:framework}).
Then, on the communication graph representation created, we apply the Node2vec~\cite{grover2016node2vec} technique to create an embedding representation for each node in the graph.
Node2vec is applied by performing random walks in the graph. 
Each random walk forms a sentence that is fed into a skip-gram model that outputs an embedding vector. 
In our case, the size of the embedding vector (i.e., the number of features) is ten, and the size of each random walk is five steps.

\item \emph{Process features.}
We also extract features describing the processes each machine used within the time window $t$. 
First, we create a multi-hot encoding matrix for the processes with $r$ (machines) rows and $p$ (processes) columns.
The value of a cell [$i$,$j$] is the number of times a machine mi used process $p_j$. 
Next, we normalized this matrix with min-max normalization for each process.
Then, to capture the distribution of a process among all machine, we computed the TFIDF value for $m_i$ and $p_j$ (by computing the document frequency value of $p_j$)~\cite{joachims1996probabilistic}.

TFIDF is a statistical measure that evaluates how relevant a word is to a document in a collection of documents. 
To match the TFIDF to our problem a word/term is replaced by a process and a document replaced by a machine. 
In addition, in order to ignore common processes \sysname uses a decay weight for TFIDF score. 
This is performed by counting the number of days each process appeared within the seven days prior to the evaluated day and updating the TFIDF score accordingly; i.e., if a process appears on multiple days its TFIDF score will be reduced: $TFIDF[i,j]=TFIDF[i,j]*0.9d$ where $d$ is the number of days which process $j$ appeared on machine $m_i$. 
The resulted matrix (Step 6 in Figure~\ref{fig:framework}) is then added it the complete feature matrix (Step 8).

\item \emph{Significant process features.}
Since typically many processes can run on each machine, in \sysname we extract features for the process with the highest TFIDF score; i.e., the process that best represents the machine. 
Using the process's unique MD5 hash value, we compute the following process-specific features: number of different PIDs, max duration for a single PID, and the average duration for all PIDs observed within the time window $t$. 
The duration of a PID is calculated using the first and last time that the process was seen within the time window $t$. 
In addition, we computed a one-hot encoding features that describes the process source directory: windows folder, program files folder, user folder, None (the process path is null), or other folder (Step 7 in Figure~\ref{fig:framework}).

\end{enumerate}

\subsection{Training A GCN-Based Autoencoder} 
In \sysname, we opted to use a GCN-based VAE for the detection of anomalous communication patterns for two main reasons.
First, since the monitored machines and the communication among them are modeled as a graph, we would naturally like to use deep learning techniques that can process and be trained with this kind of data.
Secondly, we would like to be able to consider attributes that characterize nodes and their behavior which can be applied by using GCN.
Examples of such attributes include: (1) whether the machine is an internal or external machine, (2) whether the machine is a server, workstation, or mobile computer, (3) statistical features that help in profiling the communication patterns of a machine, such as the number of communications with internal and external machines.
Unlike Node2vec or similar techniques, GCNs can utilize the structure of the graph as well as the attributes assigned to each node, and therefore, it meets our requirements.
To the best of our knowledge, we are the first to utilize a GCN-based VAE for the detection of anomalies and security incidents in communication networks.

The input to the GCN-based VAE model includes two matrices:
\textit{(1)} the normalized adjacency matrix, denoted by $CM$ (Step 3), and \textit{(2)} the features extracted for a machine which are generated by concatenating the communication-based statistical features with the node embedding and process features, denoted by $FM$ (Step 8).
In total, for each node in the node feature matrix, we extract 15 features (five statistical features and 10 embedding features). 

\begin{figure}[h]
\centering
\includegraphics[width=\linewidth]{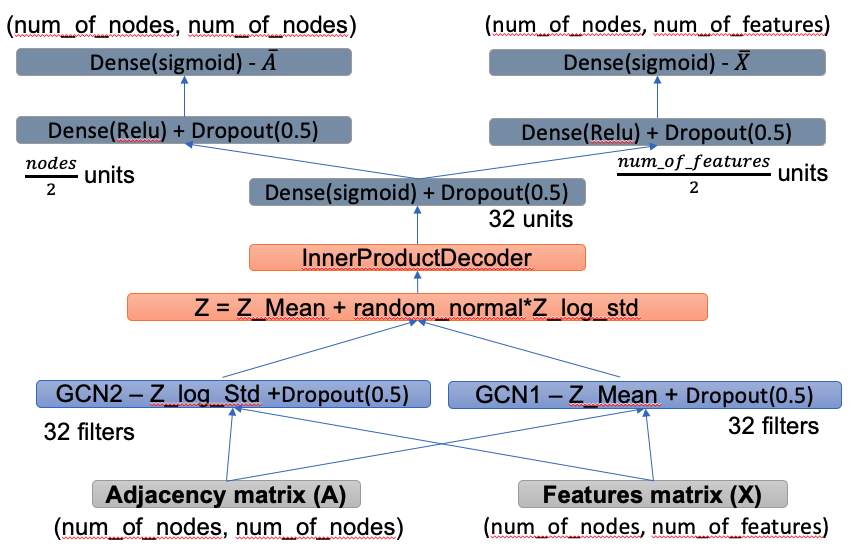}
\caption{GCN variational autoencoder architecture.}
\label{fig:GCN}
\end{figure}

\sysname uses a VAE (previously described by~\cite{an2015variational}) to reduce the input dimension and represent each node $m \in M'$  as a Gaussian distribution over the latent space (Step 9 in Figure~\ref{fig:framework}). 
The decoder tries to reconstruct the two inputs, $CM$ and $FM$, from the latent space representation.
Figure~\ref{fig:GCN} presents the architecture of the model used in this research. 
The first layers of the model are two GCNs, one for the $Z$-distribution mean and one for the $Z$-distribution standard deviation. 
Each GCN has 32 filters, and their output is connected to a dropout layer with a 50\% drop ratio. 
Next, using the two GCNs, the model creates a new latent space $Z$ that can be defined as $Z = Z_{mean} + random\_normal*Z_{std}$.
The $Z$ layer is connected to a dense layer (i.e., fully connected layer) with a sigmoid activation function which is used to represent the latent space in a normalized vector with values between 0-1.
The encoder's output is the dense layer's output, which serves as the input to an inner product decoder. 
By applying the inner product on the latent variables $Z$ and $Z^T$, the model can learn the similarity of each node within $Z$. 
Finally, the model reconstructs both matrices $CM$ and $FM$. 

The inner product decoder output connects to two layers,  one of which is in the size of $\frac{number\_of\_ features}{2}$ units dense layer for the feature matrix, and the other of which is a dense layer that is the size of $\frac{number\_of\_ nodes}{2}$. 
Then, the first dense layer is connected to a dense layer that is the size of $FM$, and the second is connected to a dense layer that is the size of the number of nodes in the graph. 
Each dense layer uses a sigmoid activation function, as \sysname tries to reconstruct the inputs and the inputs are normalized between~0-1. 

\subsection{\sysname Loss Score}
We implemented a custom loss for \sysname which divides the feature matrix into five different components according to the different subsets of features: 

\begin{itemize}
    \item Communication graph features loss $AM = MAE(CM_i - \Tilde{CM_i})$ 
    \item Communication-based statistical features loss  $SF = MAE(SF_i-\Tilde{SF_i})$
    \item Node embedding loss $EMB = MAE(EMB_i-\Tilde{EMB_i})$
    \item Processes features loss  $PO = MSE(PO_i- \Tilde{PO_i})$ where the loss of a single process is multiplied by 0.1 if the real value of process $j$ is 0 in order to handle the sparsity of the process one-hot
    \item Significant process features loss $PF = MSE(PF_i - \Tilde{PF_i})$
\end{itemize}

\noindent Finally, the reconstructed error of the model can be defined as $RE_i =AM_i + \alpha * PF_i +\beta* EMB_i + \gamma * PO_i + \delta* PF_i$ where $\alpha, \beta, \gamma, \delta$ are the weight of each features subset respectively and $\alpha + \beta + \gamma + \delta = 1$.

\subsection{\sysname Explainability}
In addition to the anomalous machine \sysname will provide explanations on why machine $m$ is an anomaly. 
In order to provide the explain features we used the original adjacency matrix $A$, original feature matrix $X$, reconstructed adjacency matrix $\Tilde{A}$, and reconstructed feature matrix $\Tilde{X}$. 
For each cell $i$,$j$ (number of communications between machine $i$ and machine $j$) in the adjacency matrix, the feature reconstructed error as the difference between the original cell in the original adjacency matrix and the reconstructed feature matrix, and can be defined as $re_{A[i,j]} = A[i,j] - \Tilde{A}[i,j]$. 

The same approach applies also for the feature matrix, and a feature reconstructed error of machine $i$ and feature $j$ can be defined as $re_{X[i,j]} = X[i,j] - \Tilde{X}[i,j]$. 
Finally, each feature with a self-reconstructed error higher than 0.2 is treated as a feature that explains the anomaly.

\subsection{Detecting Anomalous Machines}
Since we use an unsupervised approach for anomaly detection, we apply the following process for detecting the anomalous behavior of machines.
For the events logged in time window $t$ (of size $w$), we train a GCN-based VAE, as described in the previous section.
Then, we use the same data used for training the model as the test set data. , i.e., the same data used to train the model is also used as the input to the trained model after completing the training process. 

The reconstruction error of each machine in the test set serves as the machine's anomaly score. 
To calculate the machine anomaly score, \sysname calculates the difference (self-difference) in the machine's reconstruction error and multiplies it by the current reconstruction error as follows: $self-difference_i = \frac{RE_{i,t}}{\frac{\sum_{j=1}^{9}RE_{i,t-j}}{9}}$, where $i$ is the machine and $t$ is the time window. 
Finally, multiplying the new score and the reconstructed error will provide the machine's final anomaly score $final\_anomaly\_score_i = RE_i*self-difference_i$.
\sysname defines a machine as anomalous if the machine mult\_score is higher than the threshold $tr$ (Step 10 in Figure~\ref{fig:framework}).

\section{\label{sec:eval}Evaluation} 

We present the evaluation of our proposed method in two different experiments: supervised, in which we simulated malicious activity, and unsupervised, in which we applied it on real unlabeled data.

\subsection{Datasets}
For the evaluation, we used real Carbon Black EDR logs collected from a large enterprise for 16 days, from 14/10/2021 to 29/10/2021.
More specifically, we used the netconn events, which provide information on network connections that were received or initiated by an endpoint monitored by Carbon Black EDR.
In addition, for the evaluation we focused on monitoring a homogeneous subset of machines: automated teller machines (ATMs), and machine that communicated with Active Directory (AD) servers.
The first seven days (14-20) used to create the processes dataset, the following days (21-29) used to test the model performance.
On average, there were approximately 1.5 million events per day generated from approximately 3,000 ATMs, and 7 million events per day generated from 40,000 internal machines and 10 AD servers.


\subsection{Evaluation metrics}
Each alert raised by \sysname was inspected by the SOC security analyst. 
With the alert we provided the following information to facilitate the inspection process: \textit{(1)} the ID and IP address of the anomalous machine, \textit{(2)} the anomalous machine reconstructed error, and \textit{(3)} the anomalous machine final anomaly score, and \textit{(4)} a list of features with high reconstructed error as described in explain ability sub section.
The security analyst marked each alert as false positive (FP), i.e., the system should not raise this alert, or true positive (TP), i.e., the system should raise this alert and an analyst need to further investigate it.
With the assistance of the security analysts, we defined the following evaluation metrics:
\begin{itemize}
    \item \textbf{\#Anomalies} - the number of anomalous ATMs/ADs for which an alert was raised by the proposed method for a given day. 
    Since each alert needs to be manually inspected, we would like the number of anomalies to be reasonable (given the estimated amount of effort required from the security analyst, there should be less than 10 anomalies per day).
    
    \item \textbf{\%Anomalies} - the number of anomalous ATMs/ADs on a given day divided by the total number of ATMs/ADs on the same day.
    
    
    \item \textbf{\% Good Alerts} - Percentage of anomalous machines that were labeled as ``good" anomalies (i.e., should be investigated) by the security expert.
    
\end{itemize}

\subsection{Model configuration}
The model presented in the Method section was trained for 200 epochs with the Adam optimizer~\cite{kingma2014adam} and a batch size of 256.

\subsection{Results}

\noindent\textbf{Unsupervised experiment on ATMs. }
In the unsupervised experiment, we applied \sysname on the data for each day separately and identified the anomalous ATM machines for each day. 
For the ATM subset the loss hyper parameters were empirically set to be $\alpha=0.3, \beta=0.3 , \gamma=0.2, \delta=0.2$.
As mentioned, since the data is unlabeled, each anomaly identified was inspected by the security analyst.
By analyzing the model's reconstruction error, we set the decision threshold at 0.6, because it seemed to be a clear cut-off point distinguishing between normal behavior and anomalous cases.
In addition, the threshold of 0.6 provided us with a reasonable number of anomalous machines per day (less than 10).
The anomalies detected by the model can be categories into three types:
\begin{enumerate}
    \item Regular organizational endpoints (not ATMs) that were allocated with an IP address which are contained in the address range of the ATMs. 
The connection patterns of these machines are significantly different from those of ATMs, thus the high anomaly score in these cases.
\item ATMs that are in the software development phase (which we refer to as development ATMs). 
Although the SOC analyst consider these cases as false alarms, these anomalies demonstrate \sysname's ability to identify anomalous communication patterns that were not observed before.
\item Production ATMs with high anomaly scores. 
Those anomalies were caused by a technical problem with the ATM; in these cases, an IT technician ran PowerShell scripts to try to solve the problems remotely. 
Those are good anomalies, since such behavior might be performed by a remote attacker; therefore alerts should be raised as these cases require investigation by the SOC analyst to be sure the anomalous was performed by an authorized person.
\end{enumerate}

As can be seen by the results in Table~\ref{table:stats_all}, the total number of anomalous ATMs raised by \sysname is 20 in nine different days (column `Alerts'). 
65\% of those anomalies are from the first type, 15\% from the second type, and 20\% are from the third type. 
Thus, the overall good alert percentage (as marked by the security analysts) is 85\% (17 good alerts) which was stated by the security analysts as a method/system with a very high and effective precision.

\begin{table*}[h]
\centering
\begin{tabular}{|c||c|c||c|c||c|c||c|c|}
\hline
\textbf{Date} & \multicolumn{2}{c||}{\textbf{\#Events}} & \multicolumn{2}{c||}{\textbf{\#Machines}} & \multicolumn{2}{c||}{\textbf{Alerts}} & \multicolumn{2}{c|}{\textbf{Good Alerts}}\\ \hline
 & \textbf{ATM} & \textbf{AD} & \textbf{ATM} & \textbf{AD} & \textbf{ATM} & \textbf{AD} & \textbf{ATM} & \textbf{AD}\\ \hline
21/10 & 1.1M & 7.2M & 3079 & 41994 & 2 (0.065\%) & 4 (0.009\%) & 50\%  & 25\% \\ \hline
22/10 & 0.52M & 3.8M & 3312 & 44705 & 0 (0\%) & 0 (0\%) & 0\% & 0\% \\ \hline
23/10 & 1.02M & 3.1M & 3120 & 15352 & 1 (0.032\%) & 2 (0.013\%) & 50\% & 100\%  \\ \hline
24/10 & 1.17M & 3.0M & 3035 & 15874 & 2 (0.066\%) & 2 (0.012\%) & 50\% & 50\% \\ \hline
25/10 & 1.08M & 6.9M & 3003 & 34427 & 4 (0.133\%) & 1 (0.003\%) & 100\% & 100\%  \\ \hline
26/10 & 1.15M & 7.35M & 2953 & 39240 & 3 (0.101\%) & 0 (0.0\%) & 100\% & -   \\ \hline
27/10 & 1.22M & 6.7M & 3018 & 38049 & 3 (0.1\%) & 4 (0.013\%) & 33.3\% & 100\%  \\ \hline
28/10 & 1.21M & 6.85M & 3016 & 36915 & 3 (0.1\%) & 3 (0.0081\%) & 100\% & 66.7\%  \\ \hline
29/10 & 0.63M & 3.13M & 3365 & 32195 & 2 (0.059\%) & 1 (0.0031\%) & 0.0\%  & - \\ \hline

\end{tabular}%
\caption{The anomalies detected in the ATM and AD unsupervised experiments.}
\label{table:stats_all}
\end{table*}

\noindent\textbf{Unsupervised experiment on ADs. }
In this subset of machine \sysname was applied on a dataset of communications between an internal workstations or mobile devices and the organization's AD servers. 
In this dataset, the number of machines was very large (40k machines per day), which was too large to train on our servers. 
Therefore, in this case we randomly split the dataset into four groups (with approximately the same number of machines), and we applied the model  on each group separately. 
For the AD dataset the loss hyperparameters set to be $\alpha=0.4, \beta=0.2 , \gamma=0.2, \delta=0.2$, and the threshold set to be 0.018.
In this case, the anomalies detected by the model can be categorized into four groups:
(1) machines that used processes from an unknown source; (2) machines that use office-related processes for communications with an AD; (3) machines in which a change was made in the registry keys; and (4) machines that use software for setups and installs of internal applications.

As can be seen, by the results in Table~\ref{table:stats_all}, the total number of anomalous machines that communicated with an AD server raised by \sysname is 17 in nine different days (column `Alerts'). 
23.53\% of those anomalies are from the first type, 23.53\% from the second type, 11.76\% from the third type, and 41.18\% are from the fourth type. 
Thus, the overall good alert percentage (as marked by the security analysts) is 64.71\% (11 good alerts of types 1 and 4), while the alerts of types 2 and 3 are false positive.

\noindent\textbf{Supervised experiment. }
In order to test \sysname's ability to detect cyberattacks, we artificially injected connection events that simulate a cyberattack into randomly selected ATM machines. 
Specifically, we simulated the cyberattack which was reported on 2017. 
In this attack, 41 ATMs of a Taiwanese bank were attacked, and the criminals stole US\$2.5 million.\footnote{\url{https://www.europol.europa.eu/sites/default/files/documents/public_-_cashing_in_on_atm_malware.pdf}}

In our case, we can only see network-connection-related data. 
For this reason, according to the cyberattack description, we injected two types of communications: The first are communications from the infected server to an ATM machine via the internal network, and the second type are communications between the hackers and the ATMs via Wickr Me messenger app. 
Our injected logs contain 15 (eight inbound and seven outbound) communications between the infected server to an injected ATM and 15 (eight inbound and seven outbound) communications between the ATMs to the hackers via Wickr Me messenger. 
We chose 15 as the number of communications, as it is low enough to be hard to detect and high enough to allow the attackers to hack the ATMs. 
For the injection, we selected two random ATMs of the ATMs active on October 29, 2021, and injected them with the communications mentioned above.
This experiment was repeated twice, each time with different random ATMs.

In both experiments, the machines with the injected communications were detected as anomalous. On the first experiment the ATMs final anomaly score was 1.317 and 1.184.
On the second experiment the results were similar and the infected ATMs got final anomaly score 1.364 and 1.348.
Specifically, in both tests, those machines were ranked as the 1st, and 2nd most anomalous machines, whereas without the injected communications the machines were ranked 1800 and above.
This indicates the ability of \sysname to detected small anomalies.
As part of the organization's ongoing cyber defense activities (ethical hacking activity) one machine has been used to create brute force enumerations to AD servers in order to get access to internal users' accounts. 
This activity contained a total of 19 communication events between 8AM to 5PM. 
In this case, \sysname were able to detect that ethical hacking activity with a high final anomaly score of 0.0182 and the machine was ranked in the third place in~the~list.

\noindent\textbf{Model Comparison. }
To evaluate the effectiveness of \sysname in identifying anomalies, we compared the proposed model with two alternative configurations: (1) using standard autoencoder instead of variational autoencoder as presented by~\cite{ding2019deep} (with the same architecture and hyperparameters used for the VAE), and (2) training the model without the node embedding features (i.e., using only the communication graph and the statistical features).
The motivation for comparing the proposed model with these two alternatives is two-fold: First, the two alternative configuration are common in prior works and therefore can be regarded as baseline. 
Second, the two alternatives are much faster to execute and therefore more efficient and can be applied in near-real time.

The results of this experiment (using the same dataset used in the ATM unsupervised evaluation) show that the proposed model outperformed the two alternative configurations.
Specifically, the model with VAE was able to detect 82.3\% good anomalies while the model with the AE detected 65\% good anomalies.
We attribute these results to the ability of the VAE to represent the behavior of the ATMs using Gaussian distribution.
When removing the embedding features the ability of the model to detect good anomalies drops to 39\%.

\vspace{-5pt}
\section{\label{sec:related}Related Work} 

Kipf and Welling~\cite{kipf2016variational} use a GCN-based variational autoencoder as an unsupervised approach for modeling graph-based problems.
The proposed model's input includes the adjacency matrix $A$ and the feature matrix $X$, and the model attempts to reconstruct the adjacency matrix. 
Zhu et al.~\cite{zhu2020deepad} use a standard autoencoder with a GCN, and the model attempts to reconstruct both the adjacency matrix and the feature matrix. 
Similar to the approach presented by Kipf and Welling~\cite{kipf2016variational}, \sysname also uses VAE, however, in the optimization process our model is trained to reconstruct both the adjacency matrix and the feature matrix (as was done in the study by~\cite{zhu2020deepad}). 

Zhu et al.~\cite{zhu2020anomaly} present an anomaly detection approach for graphs, using GCNs and autoencoders. 
The GCNs are used as the first layer of the network and represent the graph in a compressed latent space. 
Then, the adjacency and feature matrices are reconstructed, and the reconstruction errors used to estimate the likelihood of a node being anomalous. 
In the \sysname framework, a similar approach is used to detect anomalies in the graph, but instead of using the Mean Squared Error (MSE) as the loss function, \sysname used Mean Absolute Error (MAE) is used, because we want to address the real distance between the input and the reconstructed data.



Bowman et al.~\cite{bowman2020detecting} present an unsupervised approach for computer network anomaly detection.
They use the Node2vec method for graph embedding and the embedding representation as input for anomaly detection methods. 
The authors test four methods for anomaly detection: two ML methods, local outlier factor (LOF) and isolation forest, and two rule-based methods, unknown authentication and failed login. 
In the \sysname framework, we also use Node2vec for graph embedding, but instead of applying standard ML methods, GCN is used for the anomaly detection task, as GCN is known to work well when many node's features are used as input to the model.

Other unsupervised methods represent the network traffic as a sequence of connections and feed it into an LSTM to identify anomalous sequences~\cite{bontemps2016collective,radford2018network,radford2018sequence,ahmed2016survey}. 
In this research, we focus on defining an anomalous machine and not anomalous logs or anomalous sequences, because in an unsupervised setting, the task of defining anomalous logs will result in many false alarms.


\section{Conclusions} \label{sec:conclusions}
In this paper, we presented a framework for the detection of anomalous communication patterns that could be indicative of a cyberattack on an organizational endpoint.
A GCN-based model is used to model \textit{both} the communication graphs generated for the machines in the organization as well as various features that are used to represent each machine.
The effectiveness of the proposed method was demonstrated in two experiments in which we used real, large data collected from a large organization. 
The experiments were conducted on data of ATMs and ADs logged by Carbon Black agents. 
The model performances were promising for both the supervised test (100\% success in alerting the artificially compromised ATMs and the ethical hacking machine) and the unsupervised test, 85\% of the ATM anomalies were labeled as good anomalies by the security analysts and 64.71\% of the machine that communicated with AD server was labeled as good anomaly by the security analysts.

Our assumption on the most significant success on the ATMs is because the ATMs are more homogeneous group and their behavior is very similar to each other, while machines that communicate with AD servers are different from each other as they can be different user types (admin, manager, employee, etc) and different type of machines (mobile devices, workstations, etc).

In future work, we suggest the following extensions to our framework: adapting the method so that the proposed model can be applied in near-real-time, and enhancing the explainability component of the model.

\bibliographystyle{splncs04}
\bibliography{bibliography}

\end{document}